\documentclass[journal,twoside,web]{ieeecolor}
\usepackage{generic}
\usepackage{cite}
\usepackage{amsmath,amssymb,amsfonts}
\usepackage{algorithmic}
\usepackage{graphicx}
\usepackage{textcomp}
\usepackage{booktabs}
\usepackage{makecell}
\usepackage{multirow}     
\usepackage[bookmarks=true, colorlinks=true, linkcolor=blue, citecolor=blue, urlcolor=blue]{hyperref}
\def\BibTeX{{\rm B\kern-.05em{\sc i\kern-.025em 
b}\kern-.08emT\kern-.1667em\lower.7ex\hbox{E}\kern-.125emX}}
\makeatletter
\def\ps@arxiv{%
  \def\@oddhead{}\def\@evenhead{}%
  \def\@oddfoot{\hfil\thepage\hfil}%
  \let\@evenfoot\@oddfoot}
\def\ps@titlepagestyle{\ps@arxiv}
\makeatother
\begin{document}
\title{Multi-Dimensional Entropy for Vibration Data Quality Control in Wind Turbines: Properties, Deployment, and Industrial Implications}
\author{Deshui Li\textsuperscript{\textdagger}, Xiao-Ming Yuan\textsuperscript{\textdagger,*},
Zisun Wang, and Min Zhang%
\thanks{Deshui Li, Xiao-Ming Yuan, Zisun Wang, and Min Zhang are with Mingyang Smart Energy Co., Ltd., Zhongshan, China.}
\thanks{\textsuperscript{\textdagger}Deshui Li and Xiao-Ming Yuan contributed equally to this work.}
\thanks{\raggedright\textsuperscript{*}Corresponding author: Xiao-Ming Yuan (e-mail: \nolinkurl{yuanxiaoming@mywind.com.cn}; \nolinkurl{eeyxm@outlook.com}).}
\thanks{\raggedright E-mail: \nolinkurl{lideshui@mywind.com.cn} (D. Li); \nolinkurl{wangzishun@mywind.com.cn} (Z. Wang); \nolinkurl{zhangmin01@mywind.com.cn} (M. Zhang).}
\thanks{\raggedright ORCID: D. Li, 0009-0004-8130-1880; X. Yuan, 0009-0002-0977-585X; Z. Wang, 0009-0003-1250-5356; M. Zhang, 0009-0005-2515-190X.}
}

\maketitle

\begin{abstract}
Erroneous vibration signals caused by sensor malfunction, shutdown transients, and abnormal acquisition conditions can degrade the reliability of automated industrial monitoring pipelines. This paper presents a deployment-oriented analysis of Multi-Dimensional Entropy (MDE) for vibration data quality control in wind turbines, focusing on computational efficiency, model-agnostic capability, physical interpretability, and robustness. Experiments on 57,643 labeled industrial vibration records from 12 wind farms and 14 turbine units, covering main bearings, gearboxes, and generators, together with cross-platform deployment validation and cross-turbine generalization tests on 4,152 unseen records from 3 additional wind farms, show that MDE provides a stable and discriminative feature representation across different classifiers and heterogeneous operating conditions while maintaining low computational and memory requirements. These results demonstrate that MDE can serve as a lightweight and deployment-ready feature layer for vibration data quality control, thereby improving the reliability of industrial monitoring pipelines and reducing the risk of error propagation into downstream diagnostic and prognostic tasks in wind energy applications.

\end{abstract}


\section{Introduction}
\label{sec:introduction}

Industrial monitoring systems increasingly rely on sensor data for condition assessment, fault diagnosis, and prognostic analysis. Wind turbines are a representative example, where multiple data sources, including supervisory control and data acquisition (SCADA), maintenance records, and condition monitoring systems (CMS), are integrated to support reliability-oriented operation and maintenance. SCADA data have been widely used for wind turbine condition monitoring and fault isolation, while vibration signals remain particularly important for drivetrain health assessment because they directly reflect the dynamic behavior of components such as main bearings, gearboxes, and generators~\cite{intro1,intro2}.

Meanwhile, advances in artificial intelligence and industrial informatics have accelerated the automation of diagnostic and prognostic tasks in modern industrial systems~\cite{a3,intro4}. However, this increasing reliance on monitoring data also exposes a practical but often overlooked problem: the quality of the input data is not always guaranteed. In real industrial environments, erroneous vibration signals caused by sensor malfunction, shutdown transients, abnormal acquisition conditions, or communication issues are practically unavoidable. If such signals are directly fed into automated analysis pipelines, they may lead to false alarms, missed detections, unreliable maintenance decisions, and, in severe cases, the failure to identify developing faults in time.

Although this study is conducted on wind turbine datasets, the underlying problem is not unique to wind energy. Similar data-quality-induced reliability degradation also exists in other industrial systems that depend on sensor-driven monitoring and automated decision-making. For example, high-noise environments and imperfect raw data can obscure early fault features and increase the risk of misdiagnosis or missed diagnosis in safety-critical industrial applications~\cite{intro5}. More generally, data quality has been recognized as a fundamental prerequisite for reliable data-driven analysis and decision support~\cite{b4, b5}. Therefore, vibration data quality control should be treated as an essential component of reliable industrial informatics systems rather than as a minor preprocessing step.

Multi-Dimensional Entropy (MDE) was recently introduced as a compact representation for vibration measurement validity assessment, with initial validation on heterogeneous industrial wind-turbine data~\cite{mde1}. Building on that formulation, this paper focuses on the properties required for practical deployment rather than reintroducing the method. MDE is treated as a front-end quality control layer for downstream tasks such as fault diagnosis and remaining useful life prediction, and is systematically evaluated in terms of computational efficiency, model-agnostic capability, physical interpretability, robustness, cross-turbine generalization, and cross-platform feasibility. The main contributions of this paper are summarized as follows:
\begin{itemize}
	\item Building on the original MDE formulation, a systematic deployment-oriented analysis is conducted from the perspectives of computational efficiency, model-agnostic capability, physical interpretability, robustness, cross-turbine generalization, and cross-platform feasibility.
	\item The initial three-model evidence is extended to seven classifiers, systematic feature- and dimension-level ablation, and real-data consistency analyses across sampling configurations and sensor brands.
	\item Cross-turbine generalization is evaluated on unseen datasets without retraining or fine-tuning, verifying the transferability of the proposed feature representation under practical data variations.
	\item Cross-platform experiments verify the feasibility of deploying the proposed feature layer on both centralized and edge-side hardware platforms.
\end{itemize}

\section{Related Work}

\subsection{Vibration-Based Diagnosis \& AI in Industrial Systems}

Vibration-based condition monitoring has long been a core approach for fault diagnosis in rotating machinery because vibration signals directly reflect the dynamic behavior and health condition of critical components. Representative studies have demonstrated the effectiveness of vibration-feature-based diagnosis in low-speed rotating machinery and wind turbine applications~\cite{a1,a2}.

With the rapid development of artificial intelligence, machine-learning and deep-learning methods have been widely introduced into intelligent fault diagnosis, significantly improving automation, feature learning capability, and diagnostic accuracy in industrial monitoring systems~\cite{a3,a4}. In the wind energy sector, recent reviews further show that AI-driven fault diagnosis has become an active research direction, with growing interest in lightweight models, multimodal learning, and deployment-oriented intelligent monitoring frameworks~\cite{a5}.

More recently, some studies have begun to consider nonideal sensing conditions and practical deployment constraints. For example, lightweight on-device diagnosis frameworks have been developed for noisy field measurements, highlighting the practical importance of robustness and deployment feasibility under imperfect acquisition conditions~\cite{a6}. Nevertheless, these efforts still mainly focus on improving downstream diagnostic models. In contrast, front-end quality control for erroneous vibration signals remains much less systematically investigated.

\subsection{Data Quality and Robustness in Condition Monitoring}

Data quality is a fundamental prerequisite for reliable condition monitoring and intelligent decision-making. In data-driven industrial systems, the trustworthiness of downstream diagnosis and prognostic analysis depends not only on model design, but also on the quality of the input data. This issue has been recognized in broader industrial and data-centric contexts, where data quality monitoring has been shown to be essential for maintaining the reliability of machine-learning-based applications and decision support systems~\cite{b4,b5}.

In condition monitoring, robustness has often been discussed in relation to noise, operating variability, and generalization across domains. However, the quality of the acquired input signals themselves has received much less systematic attention. Existing studies have shown that the confidence of vibration-based damage detection can be strongly affected by the quality and consistency of measured data, and dedicated data quality indicators can improve the reliability of downstream structural assessment~\cite{b6}. More recently, related work has further shown that improving the reliability of input representations can be as important as classifier design in intelligent fault diagnosis systems~\cite{b3}.

Nevertheless, front-end quality control for erroneous vibration signals remains insufficiently studied in industrial monitoring practice. In particular, there is still a lack of lightweight, interpretable, and deployment-oriented feature representations that can identify low-quality vibration data before they propagate into downstream diagnosis and prognostic tasks. This gap motivates the present study.

\subsection{Feature Engineering: Interpretability, Efficiency, and Edge Deployment}

Recent studies have increasingly emphasized interpretable, lightweight, and deployment-oriented designs for industrial condition monitoring. In particular, interpretable diagnosis frameworks have been developed using sparse filtering and algorithm-unrolling strategies to improve the physical transparency and credibility of intelligent fault diagnosis models~\cite{b7,b8}. At the same time, lightweight architectures have been proposed to reduce computational complexity and hardware requirements while maintaining competitive diagnostic accuracy in industrial environments~\cite{b10}. More recently, edge-oriented diagnosis methods, including quantized lightweight networks and edge-computing-based fault diagnosis frameworks, have further demonstrated the practical importance of low-complexity and deployable designs for industrial monitoring applications~\cite{b11,b12}.

However, these efforts are primarily centered on downstream diagnosis models. In contrast, the deployment-oriented properties of front-end feature representations, especially for vibration data quality control, have been much less systematically investigated. The initial MDE study established the seven-dimensional MDE--RMS formulation and validated it using three lightweight classifiers, controlled sampling-configuration simulations, and within-dataset visualization~\cite{mde1}. However, empirical cross-platform cost, broader cross-model behavior, dimension-level ablation, real-data subset consistency, and transfer to completely unseen turbines were not systematically examined. The present study addresses these deployment-oriented questions rather than introducing a new data-quality-control formulation.

\section{System Framework and Experimental Setup}

\subsection{Industrial Data Description}
\label{sec:industrial_data}
\subsubsection{Original Datasets}
\label{original_data}
The dataset used in this study consists of real-world vibration signals collected from operational wind turbine units across multiple wind farms. A detailed summary is provided in Table~\ref{tab:origin_dataset}. In total, the dataset covers 12 wind farms and 14 turbine units, including multiple drivetrain components, i.e., the main bearing (MB), gearbox (GB), and generator (Gene). This 57,643-record corpus is the original dataset used for initial MDE validation~\cite{mde1}; here it supports expanded cross-model, ablation, and subgroup analyses, while the unseen dataset and cross-platform measurements provide new evaluation evidence.

This dataset serves as the training and evaluation basis for the model-agnostic analysis and robustness assessment in Sections~\ref{performance_across_model} and~\ref{robustness_analysis}.

\begin{table}[htbp]
  \centering
  \caption{The Information of Original Dataset}
  \label{tab:origin_dataset}
  \setlength{\tabcolsep}{3pt}
  \renewcommand{\arraystretch}{1.2}{
  \begin{tabular}{lcccccccc}
    \toprule
    \shortstack{Model\\(MW)} & \shortstack{Wind\\Farm} & Turbine & \shortstack{Sensor \\Brand} & Unit & Position & \shortstack{Sampling \\Config\\(Hz × s)}  &\shortstack{Total\\Samples}\\
    \midrule
    3.0 &A & a & Brand 1 &  g    & MB &      12800 × 10.24 & 4646\\
    3.0 &B & b & Brand 2 & m/s$^{2}$ & MB &  12800 × 10.24 &1400\\
    3.2 &C & c & Brand 3 & m/s$^{2}$  & MB & 12800 × 10.24 &3683\\
    3.0 &D & d & Brand 4 &  g    & MB &      12800 × 10.24 &1596\\
    3.2 &E & e & Brand 5 & m/s$^{2}$  & MB & 25600 × 5.12  &2989\\
    \midrule
    3.0 &F & f & Brand 6 & m/s$^{2}$  & GB & 25600 × 5.12  &1810\\
    3.2 &C & m & Brand 3 & m/s$^{2}$  & GB & 12800 × 10.24 &3688\\
    3.2 &C & m & Brand 3 & m/s$^{2}$  & GB & 25600 × 5.12  &3683\\
    3.2 &G & g & Brand 7 &  g    & GB &      12800 × 10.24 &6130\\
    3.0 &H & h & Brand 8 &  g    & GB &  12800 × 10.24 &4684\\
    3.2 &I & i & Brand 8 &  g    & GB &  12800 × 10.24 &2018\\
    \midrule
    3.0 &F & f & Brand 6 & m/s$^{2}$  & Gene &  25600 × 5.12  &1810\\
    3.2 &J & j & Brand 1 &  g    & Gene &  25600 × 5.12  &990\\
    3.2 &K & k & Brand 6 & m/s$^{2}$  & Gene &  25600 × 5.12  &352\\
    3.2 &L & l & Brand 3 & m/s$^{2}$  & Gene &  25600 × 5.12  &8558\\
    3.2 &L & n & Brand 3 & m/s$^{2}$  & Gene &  25600 × 5.12  &9606\\
    \bottomrule
  \end{tabular}
}
\end{table}

As shown in Table~\ref{tab:origin_dataset}, the data acquisition setup exhibits significant heterogeneity in terms of sensor brands, measurement units (g and m/s$^2$), and sampling configurations (12800 Hz for 10.24 s and 25600 Hz for 5.12 s). The complete dataset contains 57,643 labeled vibration samples. All samples are manually annotated to ensure reliable ground truth.

Such diversity introduces substantial variability in signal characteristics, making the dataset representative of realistic industrial deployment conditions.

\subsubsection{Unseen Datasets}
\label{unseen_data}
To further evaluate the generalization capability of the proposed method, an unseen dataset is constructed, as summarized in Table~\ref{tab:unseen_dataset}. This dataset is exclusively used for the generalization analysis in Section~\ref{sec:unseen_data_test}. It comprises vibration signals collected from three turbine units that are completely excluded from the training dataset, ensuring a strict separation between training and testing domains.

\begin{table}[htbp]
  \centering
  \caption{The Information of Unseen Data}
  \label{tab:unseen_dataset}
  \setlength{\tabcolsep}{3pt}
  \renewcommand{\arraystretch}{1.2}{
  \begin{tabular}{ccccccccc}
    \toprule
    \shortstack{Model \\(MW)} & \shortstack{Wind \\Farm} & Turbine & \shortstack{Sensor \\ Brand} & Unit & Position & \shortstack{Sampling \\Config \\(Hz × s)}  &\shortstack{Total \\Samples}\\
    \midrule
    3 &O & o & Brand 2 & m/s$^{2}$ & MB &  12800 × 10.24 & 1077\\
    \midrule
    3.6 &P & p & Brand 2 & m/s$^{2}$ & GB &  25600 × 5.12  &1030\\
    \midrule
    3.2 &Q & q & Brand 6 & m/s$^{2}$ & Gene &  25600 × 5.12 &908 \\
    3.6 &P & p & Brand 2 & m/s$^{2}$ & Gene &  25600 × 5.12 &1137\\
    \bottomrule
  \end{tabular}
}
\end{table}

The unseen dataset preserves the characteristics of real-world monitoring data, including different components — main bearing (MB), gearbox (GB), and generator (Gene) — and heterogeneous acquisition settings. All samples are manually labeled; however, minor labeling uncertainty may exist due to practical constraints in industrial annotation.

\subsection{Deployment Architecture Validation}
To validate the practical applicability of the proposed MDE method under realistic wind farm condition monitoring scenarios, two representative deployment architectures are considered: a centralized architecture and an edge-preprocessing architecture.

\subsubsection{Overall Architecture and Data Flow}
The \textbf{centralized architecture} follows a conventional paradigm in which raw vibration signals are transmitted to a central server for MDE-based feature extraction and subsequent quality assessment. This approach leverages abundant computational resources and is suitable for large-scale data processing.

In contrast, the \textbf{edge-preprocessing architecture} performs MDE feature extraction and preliminary data quality control directly at the edge. By reducing the volume of transmitted data, this architecture improves system responsiveness and enhances robustness under limited or unstable network conditions. It is particularly suitable for scenarios where continuous connectivity to a central server cannot be guaranteed.

\subsubsection{Centralized Server Deployment}
In the centralized deployment scenario, a high-performance workstation is used to emulate the wind farm data center environment. This configuration is responsible for database management, MDE feature extraction, data quality assessment, and batch classification. It represents a typical high-compute-capacity node capable of supporting large-scale storage and real-time analysis tasks. The hardware specifications are provided in Table~\ref{tab:hardware_platforms}.

\subsubsection{Industrial Edge Deployment}
For industrial edge deployment, an N100 industrial PC is adopted as the edge computing node. This fanless DIN-rail system is designed for harsh deployment environments, such as wind turbine tower bases or nacelles. Its hardware specifications are summarized in Table~\ref{tab:hardware_platforms}.

\subsubsection{Ultra-Low-Cost Edge Deployment}
To further evaluate deployment feasibility under resource-constrained conditions, the proposed pipeline is implemented on a Jetson Orin Nano platform. This setup represents a low-cost edge computing scenario, and its specifications are listed in Table~\ref{tab:hardware_platforms}.

By executing MDE feature extraction and classification locally, the system reduces reliance on continuous central server support, enabling a lightweight and cost-effective solution for distributed vibration data quality control.

\subsection{Feature Construction (MDE \& RMS)}

\label{sec:feature_construction}

Following the original MDE formulation~\cite{mde1}, each vibration record is described using entropy measures computed from different signal perspectives, where each perspective corresponds to one physical dimension of signal characterization. In random vibration analysis, vibration signals can be described not only by deterministic waveform or spectral amplitudes, but also through probabilistic representations such as probability density functions~\cite{s2}. MDE maps signal-related quantities from multiple perspectives into discrete probability distributions, including time-domain amplitude distributions, spectral-amplitude distributions, and frequency-band energy distributions. The distributional characteristics are then quantified following the Shannon entropy principle~\cite{s1}. Given a normalized distribution $\mathbf{p}=\{p_i\}_{i=1}^{K}$, the entropy is computed as
\[
H(\mathbf{p})=-\sum_{i=1}^{K} p_i \log_2(p_i+\epsilon),
\]
where $K$ denotes the number of bins or frequency bands, and $\epsilon$ is a small positive constant used to avoid numerical singularity.

Based on this formulation, three entropy-based descriptions are considered. Time-amplitude entropy describes the distribution of waveform amplitudes in the time domain, and its absolute-value variant further characterizes the amplitude distribution after removing sign information. Spectrum-amplitude entropy describes the distribution of spectral amplitudes obtained from the single-sided spectrum. Frequency-band energy entropy describes the allocation of spectral energy across predefined frequency bands. In addition, RMS is included as a compact energy descriptor to reflect the global vibration amplitude level.

Therefore, MDE and RMS jointly characterize vibration data quality from four complementary dimensions: energy level, time-domain amplitude distribution, spectral organization, and band-wise energy allocation. The specific parameterization used to instantiate these descriptors into the final feature vector is described in Section~\ref{sec:mde_parameter_settings}.

\subsection{Classification Module Design}
\label{classification_module}
To demonstrate the model-agnostic capability of the proposed MDE feature set, seven classifiers from four fundamentally different modeling paradigms are selected: Logistic Regression and Linear SVM (linear models), Kernel SVM with RBF kernel (nonlinear kernel), Decision Tree, Random Forest, and LightGBM (tree-based ensemble), and a shallow Multi-Layer Perceptron (neural network). This diverse selection verifies that MDE provides discriminative features regardless of the underlying classification mechanism.

\subsection{Evaluation Metrics}
\label{evaluation metrics}
Five standard classification metrics are employed: Accuracy (ACC), True Detection Rate (TDR), False Positive Rate (FPR), F1-score, and Area Under the ROC Curve (AUC). Their definitions are given as follows:
\begin{itemize}
\item $\mathrm{ACC}=(TP+TN)/(TP+TN+FP+FN)$
\item $\mathrm{TDR}=TP/(TP+FN)$
\item $\mathrm{FPR}=FP/(FP+TN)$
\item $\mathrm{F1}=2 \times \mathrm{Precision} \times \mathrm{TDR} / (\mathrm{Precision}+\mathrm{TDR})$, where $\mathrm{Precision}=TP/(TP+FP)$
\item AUC evaluates the separability of the classifier across all decision thresholds.
\end{itemize}
Here, TP, TN, FP, and FN denote true positives, true negatives, false positives, and false negatives, respectively, where the positive class refers to erroneous vibration signals.

\subsection{Experimental Settings}
\subsubsection{MDE Parameter Settings}
\label{sec:mde_parameter_settings}

Based on the feature construction described in Section~\ref{sec:feature_construction}, the MDE descriptors are instantiated into a seven-dimensional feature vector:
\[
\mathbf{x} =
[\mathrm{RMS}, \mathrm{TAE}, \mathrm{ATAE}, \mathrm{SpAE}_1, \mathrm{SpAE}_2, \mathrm{FBEE}_1, \mathrm{FBEE}_2].
\]
Before feature extraction, vibration signals measured in $g$ are converted to m/s$^2$, and the mean value of each record is removed. RMS is then calculated from the detrended signal as the energy descriptor. The single-sided amplitude spectrum is obtained using FFT with amplitude normalization.

The parameter settings used to construct the MDE features are summarized in Table~\ref{tab:mde_parameters}. TAE and ATAE are computed from the amplitude distributions of the detrended signal and its absolute-valued form, respectively, using a fixed amplitude bin width of $1/100$. SpAE$_1$ is computed from the spectral-amplitude distribution using a fixed bin width of $1/1000$, whereas SpAE$_2$ uses an adaptive bin width determined by the RMS of the corresponding signal. FBEE$_1$ divides the spectrum into two regions using 500 Hz as the threshold frequency, while FBEE$_2$ partitions the spectrum into multiple frequency bands with a fixed interval of 500 Hz.

\begin{table}[htbp]
\centering
\footnotesize
\caption{MDE Parameter Configuration}
\label{tab:mde_parameters}
\begin{tabular}{lll}
\toprule
Feature & Signal representation & Parameter setting \\
\midrule
RMS & Time-domain signal & Detrended signal \\
TAE & Time-domain amplitude & Bin width = $1/100$ \\
ATAE & Abs. time-domain amplitude & Bin width = $1/100$ \\
SpAE$_1$ & Spectral amplitude & Bin width = $1/1000$ \\
SpAE$_2$ & Spectral amplitude & Adaptive bin width = RMS$/5$ \\
FBEE$_1$ & Band energy & Threshold = 500 Hz \\
FBEE$_2$ & Band energy & Band interval = 500 Hz \\
\bottomrule
\end{tabular}
\end{table}

The same parameter configuration is used across all classifiers, sampling configurations, sensor brands, unseen datasets, and hardware platforms. No dataset-specific or platform-specific parameter tuning is performed. Therefore, the reported results mainly reflect the intrinsic robustness and transferability of the MDE representation rather than adaptation to a particular data subset or deployment environment.

\subsubsection{Classifier Hyperparameter Settings}
The hyperparameters of all classifiers described in Section~\ref{classification_module} are listed in Table~\ref{tab:hyperparameters}.
\begin{table}[htbp]
\centering
\caption{Hyperparameter Configuration of Classifiers}
\label{tab:hyperparameters}
\resizebox{\columnwidth}{!}{%
\begin{tabular}{ll}
\toprule
\textbf{Model} & \textbf{Hyperparameters} \\
\midrule
LR & penalty=l2, C=1.0, solver=lbfgs, max\_iter=1000 \\
\midrule
Linear SVM & kernel=linear, C=0.01, probability=True \\
\midrule
Kernel SVM (RBF) & kernel=rbf, C=100, gamma=1, probability=True \\
\midrule
Decision Tree & max\_depth=5, min\_samples\_split=10 \\
\midrule
Random Forest & \makecell[l]{n\_estimators=200, max\_depth=30\\ min\_samples\_split=5, n\_jobs=-1} \\
\midrule
LightGBM & \makecell[l]{num\_leaves=127, learning\_rate=0.05\\ max\_depth=10, n\_estimators=100} \\
\midrule
MLP & \makecell[l]{hidden\_layers=(64,32), activation=relu, solver=adam,\\ max\_iter=1000, early\_stopping=True} \\
\bottomrule
\end{tabular}
}
\end{table}

\subsubsection{Hardware Platform}
The deployment experiments were conducted on three hardware platforms, as summarized in Table~\ref{tab:hardware_platforms}.

\begin{table}[htbp]
\centering
\caption{Hardware Platforms for Deployment Evaluation}
\label{tab:hardware_platforms}
\resizebox{\columnwidth}{!}{%
\begin{tabular}{cccc}
\toprule
\textbf{Platform} & \textbf{CPU}  & \textbf{RAM} & \makecell{\textbf{Software}\\ \textbf{Environment}} \\
\midrule
WinPC     & Intel Core i5-12400F & 32 GB & \makecell{Windows 10 \\Python 3.10.13} \\
\midrule
N100 IPC  & Intel N100    & 4 GB  & \makecell{Ubuntu 22.04.5\\ Python 3.10.12}  \\
\midrule
\makecell{Jetson\\Orin\\ Nano} & ARM Cortex-A78AE & 8 GB & \makecell{JetPack 6.0\\ Ubuntu 22.04\\ Python 3.10.12}  \\
\bottomrule
\end{tabular}
}
\end{table}

\subsubsection{Theoretical Complexity}
The MDE feature extraction involves computing seven entropy dimensions from a single $d$-point vibration signal. The dominant computational cost stems from the fast Fourier transform (FFT) and histogram-based entropy calculations. The FFT step requires $O(d \log d)$ operations, while each of the histogram-based entropies operates in $O(d)$ time after binning. The overall time complexity of MDE extraction is therefore $O(d \log d)$, and the space complexity is $O(d)$ for storing the signal, FFT coefficients, and intermediate histograms. The model inference step adds a marginal constant cost for single-sample prediction. This low-order polynomial complexity ensures that the pipeline scales gracefully with signal length and remains suitable for resource-constrained edge deployment.
\subsubsection{Computational Cost Measurement}

The end-to-end processing cost of the MDE pipeline was measured under a sequential multi-signal scheme. For each data scale $n \in \{1, 50, 100, 500, 1000\}$, a total of $n$ distinct vibration signals were randomly drawn from the unseen cross-turbine dataset. Each signal was independently processed through the full pipeline: raw data loading from local storage, seven-dimensional MDE feature computation, and model inference. The total execution time $T_{\text{total}}$ was recorded as the wall-clock duration from the initiation of the first signal to the completion of the $n$-th signal, and the average per-signal processing time was computed as $T_{\text{avg}} = T_{\text{total}} / n$. Throughout this process, the peak physical memory increment was tracked using process-level memory monitoring.

\section{Computational Efficiency and Cross-Platform Evaluation}



\subsection{Runtime and Memory Usage Analysis}

In this section, we evaluate the end-to-end efficiency of the proposed MDE-based framework on three different platforms: a central server (high-performance PC), an industrial PC (IPC), and a low-power edge device (Jetson Orin Nano 8GB). The complete pipeline --- .txt file reading, MDE feature extraction, and model inference --- was tested under five data scales (1, 50, 100, 500, and 1000 samples). Due to page limitations, the main text reports four representative classifiers and three data scales (1, 100, and 1000), covering linear, kernel-based, tree-based, and neural paradigms. Results for all seven classifiers and all five data scales are provided in the supplementary material.

Table~\ref{tab:runtime_memory_comparison} summarizes the runtime and peak memory usage of the four representative classifiers on the three platforms.

\begin{table}[htbp]
    \centering
    \footnotesize
    \caption{Runtime and memory usage comparison across platforms}
    \label{tab:runtime_memory_comparison}
    \begin{tabular}{llcccc}
    \toprule
    Model & Platform & Data & Total  & Avg Time   &Peak  \\
          &          & Size & Time   & per Sample &Memory\\
          &          &      & (s)    & (ms)       &(MB)  \\
    \midrule
    \multirow{9}{*}{\textbf{LR}}     
        & Win PC    & 1    & 0.046  & 45.653  & 2.8008\\
        & Win PC    & 100  & 4.330  & 43.302  & 1.3555\\
        & Win PC    & 1000 & 41.508 & 41.508  & 4.5117\\
        \cmidrule(lr){2-6}
        & N100 IPC  & 1    & 0.028  & 28.131   & 0 \\
        & N100 IPC  & 100  & 2.560  & 25.597  & 3 \\
        & N100 IPC  & 1000 & 25.830 & 25.830  & 3 \\
        \cmidrule(lr){2-6}
        & Orin Nano & 1    & 0.061 & 60.553 & 0 \\
        & Orin Nano & 100  & 5.534 & 55.337 & 0 \\
        & Orin Nano & 1000 & 54.832 & 54.832 & 0 \\
    \midrule
    \multirow{9}{*}{\begin{tabular}{@{}l@{}}\textbf{Kernel}\\ \textbf{SVM}\\  \textbf{(RBF)}\end{tabular}}
        & Win PC    & 1    & 0.045   & 44.975  & 3.6172 \\
        & Win PC    & 100  & 4.402   & 44.02   & 1.3555 \\
        & Win PC    & 1000 & 41.548  & 41.548  & 4.55859 \\
        \cmidrule(lr){2-6}
        & N100 IPC  & 1    & 0.029 & 28.984  & 0.2578\\
        & N100 IPC  & 100  & 2.571 & 25.708  & 3 \\
        & N100 IPC  & 1000 & 25.923 & 25.923  & 3 \\
        \cmidrule(lr){2-6}
        & Orin Nano & 1    & 0.077  & 77.278& 0 \\
        & Orin Nano & 100  & 5.572  & 55.724 & 0 \\
        & Orin Nano & 1000 & 54.691 & 54.691 & 0 \\                  
    \midrule
    \multirow{9}{*}{\begin{tabular}{@{}l@{}}\textbf{Decision}\\ \textbf{Tree}\end{tabular}}
        & Win PC    & 1    & 0.065  & 64.553  & 2.3711 \\
        & Win PC    & 100  & 5.875  & 58.748    & 1.3555 \\
        & Win PC    & 1000 &  47.050  & 47.050   & 4.9648 \\
        \cmidrule(lr){2-6}
        & N100 IPC  & 1    & 0.029  & 29.431  & 0.0586\\
        & N100 IPC  & 100  & 2.570  & 25.701  & 3 \\
        & N100 IPC  & 1000 & 25.893 & 25.893  & 3 \\
        \cmidrule(lr){2-6}
        & Orin Nano & 1    & 0.061 & 60.883 & 0.0039\\
        & Orin Nano & 100  & 5.536 & 55.359 & 0 \\
        & Orin Nano & 1000 & 54.386 & 54.386 & 0 \\
    \midrule
    \multirow{9}{*}{\textbf{MLP}}                    
        & Win PC    & 1    & 0.048 & 47.743  & 2.7305 \\
        & Win PC    & 100  & 4.422 & 44.219   & 1.3555 \\
        & Win PC    & 1000 & 40.878  & 40.878   & 5.4141 \\
        \cmidrule(lr){2-6}
        & N100 IPC  & 1    & 0.028 & 28.131 & 0.1758\\
        & N100 IPC  & 100  & 2.568 & 25.683 & 3 \\
        & N100 IPC  & 1000 & 25.865 & 25.865 & 3 \\
        \cmidrule(lr){2-6}
        & Orin Nano & 1    & 0.061  & 60.812 & 0 \\
        & Orin Nano & 100  & 5.555  & 55.548 & 0 \\
        & Orin Nano & 1000 & 54.453 & 54.453 & 0 \\
    \bottomrule
    \end{tabular}
\end{table}

\subsection{Deployment Perspective}
\label{sec:deployment_perspective}
Several observations emerge from Table~\ref{tab:runtime_memory_comparison}. First, the N100 IPC achieved a per-signal processing time of 25.68--29.43~ms, compared with 40.88--64.55~ms on the Win PC and 54.39--77.28~ms on the Jetson Orin Nano, representing relative reductions of 37.2\%--54.4\% and 52.8\%--61.9\%, respectively. This counter-intuitive result is attributable to the lighter system overhead of the Linux operating system and the stronger single-core compute capability of the N100 relative to the ARM Cortex-A78AE cores in the Jetson, indicating that both OS efficiency and CPU microarchitecture can outweigh raw hardware specifications for lightweight inference workloads. Second, the average per-signal time decreased noticeably from scale $n{=}1$ to $n{=}100$ but plateaued between $n{=}100$ and $n{=}1000$ on certain platform--model combinations, confirming that fixed initialization overheads are amortized at moderate scales while the marginal per-signal cost remains stable. Third, peak memory increments remained below 5.5~MB across all platforms and models, well within the 4~GB capacity of the lowest-specification device. Fourth, per-signal times were comparable across classifiers on each platform, reflecting that the dominant computational cost lies in MDE feature extraction rather than model inference. In practical wind farm operation, vibration data are collected only 4--8 times per measurement point per day with each recording lasting 5 or 10~s; the observed processing latency of 25--64~ms per recording is therefore negligible, confirming the pipeline's suitability for real-time quality screening on resource-constrained edge hardware.







\section{Comprehensive Performance Evaluation}

\subsection{Performance Across Models}
\label{performance_across_model}
To extend the initial three-model validation~\cite{mde1}, ten independent repeated experiments are conducted for each of seven classifiers. All in-domain evaluations use stratified random record-level splitting after pooling the available records, without turbine- or wind-farm-level grouping. This protocol reflects routine screening within an established fleet, where historical records from multiple installed units are jointly available; cross-turbine transfer is assessed separately on the unseen dataset in Section~\ref{sec:unseen_data_test}. In each run, only the split seed is varied while all other configurations remain unchanged. The averaged results are reported in Table~\ref{tab:performance-across-models}.

\begin{table}[htbp]
\centering
\footnotesize
\caption{Summary of average evaluation metrics across 10 independent runs}
\label{tab:performance-across-models}
\setlength{\tabcolsep}{2.5pt}
\begin{tabular}{lccccc}
\toprule
Model                  & ACC (\%)   & TDR (\%)   & FPR (\%)   & F1-Score (\%)     & AUC\\
\midrule
LR                     & $98.79$    & $98.99$    & $1.85$     & $99.20$    & $0.9961$ \\
Linear SVM             & $98.79$    & $98.96$    & $1.75$     & $99.20$    & $0.9955$ \\
Kernel SVM (RBF)       & $99.25$    & $99.36$    & $1.12$     & $99.50$    & $0.9950$ \\
Decision Tree          & $99.12$    & $99.35$    & $1.61$     & $99.42$    & $0.9961$ \\
Random Forest          & $99.25$    & $99.37$    & $1.13$     & $99.50$    & $0.9978$ \\
LightGBM               & $99.24$    & $99.36$    & $1.13$     & $99.50$    & $0.9979$ \\
MLP                    & $99.07$    & $99.29$    & $1.62$     & $99.39$    & $0.9970$ \\
\bottomrule
\end{tabular}
\end{table}

As observed, all classifiers consistently achieve high performance (ACC $>$ 98.7\%, AUC $\approx$ 0.99), with only marginal variation across different model structures. This indicates that the discriminative power of the MDE feature representation is largely independent of the downstream classifier, confirming its strong model-agnostic property.

\subsection{Robustness Analysis}
\label{robustness_analysis}
The robustness of the proposed MDE feature representation is evaluated through parameter sensitivity and performance consistency across the sampling configurations and sensor brands represented in the dataset. The objective is to assess the stability and reliability of the extracted features under variations in data acquisition conditions and system configurations. In this subsection, LightGBM is adopted as the unified classifier to ensure consistency in performance comparison.

\subsubsection{Feature and Dimension Sensitivity}

From the perspective of MDE construction, the extracted features can be categorized into four dimensions: 
\begin{itemize}
    \item \textbf{Energy dimension}: represented by RMS.
    \item \textbf{Time-domain entropy dimension}: represented by TAE and ATAE.
    \item \textbf{Spectral amplitude entropy dimension}: represented by SpAE$_1$ and SpAE$_2$.
    \item \textbf{Frequency band energy entropy dimension}: represented by FBEE$_1$ and FBEE$_2$.
\end{itemize}

To comprehensively analyze parameter sensitivity, experiments are conducted from two complementary aspects, namely feature-level analysis and dimension-level analysis.

\paragraph{Feature-level Analysis}

To evaluate the intrinsic discriminative capability of each entropy component, a single-feature analysis is performed. The results are summarized in Table~\ref{tab:performance-single-feature}.

\begin{table}[htbp]
\centering
\footnotesize
\caption{Summary of average evaluation metrics across 10 independent runs in single Feature}
\label{tab:performance-single-feature}
\setlength{\tabcolsep}{2.5pt}
\begin{tabular}{lccccc}
\toprule
Feature          & ACC (\%)   & TDR (\%)   & FPR (\%)   & F1-Score (\%)     & AUC\\
\midrule
TAE              & $98.13$    & $98.14$    & $1.88$     & $98.76$    & $0.9943$ \\
ATAE             & $97.98$    & $97.83$    & $1.57$     & $98.66$    & $0.9939$ \\
SpAE$_1$         & $97.33$    & $98.47$    & $6.23$     & $98.25$    & $0.9946$ \\
SpAE$_2$         & $76.08$    & $99.68$    & $98.05$    & $86.34$    & $0.5869$ \\
FBEE$_1$         & $79.22$    & $96.03$    & $7.36$     & $87.52$    & $0.7786$ \\
FBEE$_2$         & $89.27$    & $97.91$    & $37.89$    & $93.26$    & $0.8716$ \\
RMS              & $97.87$    & $97.82$    & $1.99$     & $98.58$    & $0.9928$ \\
\bottomrule
\end{tabular}
\end{table}

It can be observed that TAE, ATAE, SpAE$_1$, and RMS achieve consistently high classification performance across all evaluation metrics, indicating strong individual discriminative capability. In contrast, SpAE$_2$ and the band energy features (FBEE$_1$ and FBEE$_2$) exhibit relatively weaker standalone performance, particularly reflected by elevated false positive rates.

However, it is important to note that these weaker features are not redundant in a strict sense. Instead, they provide complementary information that becomes effective when combined with other features, which will be further validated in the subsequent dimension-level analysis.

\paragraph{Dimension-level Analysis}

To further investigate the effect of feature aggregation, combinations of different feature dimensions are evaluated, as summarized in Table~\ref{tab:performance-dimension-level}. For clarity, the dimension--feature correspondence is defined as: Energy (RMS), Time (TAE, ATAE), Spectrum (SpAE$_1$, SpAE$_2$), and Band (FBEE$_1$, FBEE$_1$).

\begin{table}[htbp]
\centering
\footnotesize
\caption{Summary of average evaluation metrics across 10 independent runs in dimension level}
\label{tab:performance-dimension-level}
\setlength{\tabcolsep}{2.5pt}
\begin{tabular}{lccccc}
\toprule
Case                  & ACC (\%)   & TDR (\%)   & FPR (\%)   & F1-Score (\%)     & AUC\\
\midrule
Full MDE              & $99.24$    & $99.36$    & $1.13$     & $99.50$    & $0.9979$ \\
\midrule
w/o RMS               & $99.24$    & $99.36$    & $1.13$     & $99.50$    & $0.9978$ \\
w/o Time              & $99.24$    & $99.36$    & $1.13$     & $99.50$    & $0.9977$ \\
w/o Spectrum          & $99.17$    & $99.32$    & $1.28$     & $99.45$    & $0.9977$ \\
w/o Band              & $98.96$    & $99.25$    & $1.95$     & $99.32$    & $0.9971$ \\
\midrule
Band+RMS              & $99.12$    & $99.30$    & $1.45$     & $99.42$    & $0.9975$ \\
Spectrum+Band         & $99.15$    & $99.30$    & $1.33$     & $99.44$    & $0.9976$ \\
Spectrum+RMS          & $98.62$    & $98.98$    & $2.51$     & $99.09$    & $0.9966$ \\
Time+Band             & $99.15$    & $99.29$    & $1.31$     & $99.44$    & $0.9975$ \\
Time+RMS              & $98.77$    & $98.88$    & $1.59$     & $99.18$    & $0.9961$ \\
Time+Spectrum         & $98.84$    & $99.15$    & $2.13$     & $99.24$    & $0.9969$ \\
\midrule
only RMS              & $97.87$    & $97.82$    & $1.99$     & $98.58$    & $0.9928$ \\
only Time             & $98.57$    & $98.74$    & $1.97$     & $99.05$    & $0.9954$ \\
only Spectrum         & $98.22$    & $98.64$    & $3.08$     & $98.83$    & $0.9955$ \\
only Band             & $95.01$    & $97.24$    & $12$       & $96.73$    & $0.9774$ \\
\bottomrule
\end{tabular}
\end{table}

Several important observations can be drawn. First, high classification performance can already be achieved using a single dimension, particularly the time and spectral dimensions, which is consistent with the feature-level findings. Second, incorporating additional dimensions leads to performance improvement; however, the gain rapidly diminishes as more dimensions are included. Specifically, the performance difference between three-dimensional and full four-dimensional configurations is marginal.

This behavior indicates that most discriminative information is captured by a subset of dimensions, while the remaining components introduce a certain degree of redundancy. Importantly, this redundancy is beneficial rather than detrimental, as it enhances robustness by providing alternative representations of the underlying signal characteristics.

Overall, the results demonstrate that the proposed MDE representation achieves a favorable balance between discriminability and redundancy. Even under reduced feature conditions, the model maintains stable performance, confirming its robustness against feature degradation and incomplete information scenarios.

\paragraph{Redundancy and Robustness Discussion}
By jointly analyzing the feature-level and dimension-level results, it can be concluded that the robustness of the proposed MDE representation primarily stems from its multi-dimensional redundant design. Strong individual features ensure baseline performance, while complementary weaker features enhance stability under varying conditions.

\subsubsection{Performance Consistency Across Sampling Configurations}
Whereas the initial study examined sampling invariance using controlled synthetic signals~\cite{mde1}, the present analysis evaluates empirical consistency in industrial data. The dataset is partitioned into two subsets corresponding to 12800 Hz (10.24 s) and 25600 Hz (5.12 s), as summarized in Table~\ref{tab:origin_dataset}. For each subset, 10 independent experiments are conducted using different random seeds, while keeping all other settings unchanged.

The averaged performance metrics are reported in Table~\ref{tab:Sampling Configuration Robustness}. It can be observed that the classification performance remains consistently high across different sampling configurations, with negligible variation in ACC, F1-score, and AUC.

\begin{table}[htbp]
\centering
\footnotesize
\caption{Summary of average evaluation metrics across sampling configurations and 10 independent runs}
\label{tab:Sampling Configuration Robustness}
\setlength{\tabcolsep}{2.5pt}
\begin{tabular}{lcccccc}
\toprule
\shortstack{Samp. \\Config. } & Model & \shortstack{ACC \\(\%)}  & \shortstack{TDR \\(\%)}  & \shortstack{FPR \\(\%)}   & \shortstack{F1-Score \\(\%)}     & AUC\\
\midrule
  \multirow{7}{*}{\textbf{\shortstack{12800Hz\\10.24s}}}   
& LR                     & $98.80$    & $99.00$    & $1.81$     & $99.20$    & $0.9975$ \\
& Linear SVM             & $98.67$    & $98.87$    & $1.93$     & $99.11$    & $0.9946$ \\
& RBF SVM                & $99.57$    & $99.67$    & $0.73$     & $99.71$    & $0.9969$ \\
& Decision Tree          & $99.43$    & $99.64$    & $1.19$     & $99.62$    & $0.9970$ \\
& Random Forest          & $99.50$    & $99.64$    & $0.90$     & $99.67$    & $0.9990$ \\
& LightGBM               & $99.49$    & $99.64$    & $1.19$     & $99.62$    & $0.9970$ \\
& MLP                    & $99.34$    & $99.54$    & $1.25$     & $99.56$    & $0.9986$ \\
\midrule 
  \multirow{7}{*}{\textbf{\shortstack{25600Hz\\5.12s}}}
& LR                     & $99.01$    & $99.25$    & $1.78$     & $99.35$    & $0.9965$ \\
& Linear SVM             & $99.01$    & $99.22$    & $1.67$     & $99.35$    & $0.9966$ \\
& RBF SVM                & $99.22$    & $99.23$    & $0.79$     & $99.49$    & $0.9940$ \\
& Decision Tree          & $99.11$    & $99.13$    & $0.95$     & $99.42$    & $0.9948$ \\
& Random Forest          & $99.23$    & $99.24$    & $0.79$     & $99.50$    & $0.9967$ \\
& LightGBM               & $99.22$    & $99.20$    & $0.74$     & $99.49$    & $0.9967$ \\
& MLP                    & $98.99$    & $99.21$    & $1.73$     & $99.34$    & $0.9965$ \\
\bottomrule
\end{tabular}
\end{table}

These results show consistent performance across the two acquisition configurations represented in the dataset. Since training and testing are performed within each configuration rather than across an unseen configuration, this analysis evaluates configuration-wise consistency rather than cross-configuration transfer.

\subsubsection{Performance Consistency Across Sensor Brands}
To evaluate robustness against sensor heterogeneity, the dataset is further divided into eight subsets according to different sensor brands. For each subset, 10 independent experiments are conducted across seven classifiers, and the performance is averaged over both models and runs.

The results are summarized in Table~\ref{tab:performance-across-bands}. Overall, the proposed method achieves consistently high performance across different sensor brands, with ACC and F1-score remaining above 98\% in most cases. Although a relatively higher FPR is observed for Brand 4, the corresponding ACC and AUC remain competitive compared to other sensors. This suggests that the impact of sensor-specific characteristics on the proposed feature representation is limited.

\begin{table}[htbp]
\centering
\footnotesize
\caption{Summary of average evaluation metrics across sensor brands and 10 independent runs}
\label{tab:performance-across-bands}
\setlength{\tabcolsep}{2.5pt}
\begin{tabular}{lccccc}
\toprule
Sensor Brand           & ACC (\%)   & TDR (\%)   & FPR (\%)   & F1-Score (\%)     & AUC\\
\midrule
Brand 1                & $99.45$    & $99.62$    & $2.19$     & $99.70$    & $0.9966$ \\
Brand 2                & $99.71$    & $99.86$    & $7.75$     & $99.85$    & $0.9983$ \\
Brand 3                & $99.12$    & $99.29$    & $1.41$     & $99.42$    & $0.9961$ \\
Brand 4                & $98.30$    & $99.65$    & $28.65$    & $99.11$    & $0.9567$ \\
Brand 5                & $98.79$    & $98.80$    & $1.12$     & $99.13$    & $0.9956$ \\
Brand 6                & $98.86$    & $98.71$    & $0.85$     & $99.14$    & $0.9938$ \\
Brand 7                & $99.77$    & $99.80$    & $0.32$     & $99.85$    & $0.9987$ \\
Brand 8                & $98.81$    & $99.78$    & $0.13$     & $99.85$    & $0.9993$ \\
\bottomrule
\end{tabular}
\end{table}

These results demonstrate performance consistency across the sensor-brand subsets represented in the dataset. The experiment does not target transfer to an unseen sensor brand; instead, it evaluates whether MDE remains effective under the heterogeneous sensor deployments encountered in the available industrial data.

\subsection{Generalization Analysis}
\label{sec:unseen_data_test}

The unseen dataset used for generalization evaluation is described in Section~\ref{sec:industrial_data}. In this experiment, the trained models obtained from the original dataset are directly applied to the unseen dataset without any retraining or fine-tuning. The evaluation results are presented in Table~\ref{tab:generalization-analysis-unseen-data}.

\begin{table}[htbp]
\centering
\footnotesize
\caption{Generalization Test Evaluation Metric Under Unseen Data}
\label{tab:generalization-analysis-unseen-data}
\setlength{\tabcolsep}{2.5pt}
\begin{tabular}{lccccc}
\toprule
Model                  & ACC (\%)   & TDR (\%)   & FPR (\%)   & F1-Score (\%) & AUC               \\
\midrule
LR                     & $99.23$    & $99.28$    & $0.95$     & $99.50$    & $0.9975$ \\
Linear SVM             & $99.13$    & $99.19$    & $1.05$     & $99.44$    & $0.9964$ \\
Kernel SVM (RBF)       & $99.21$    & $99.22$    & $0.84$     & $99.48$    & $0.9927$ \\
Decision Tree          & $99.42$    & $99.38$    & $0.42$     & $99.62$    & $0.9958$ \\
Random Forest          & $99.40$    & $99.38$    & $0.53$     & $99.61$    & $0.9958$ \\
LightGBM               & $99.30$    & $99.25$    & $0.53$     & $99.55$    & $0.9968$ \\
MLP                    & $99.45$    & $99.41$    & $0.42$     & $99.64$    & $0.9980$ \\
\bottomrule
\end{tabular}
\end{table}

It can be observed that all models maintain consistently high performance on unseen data, with ACC exceeding 99\% and AUC remaining close to 1.0. This demonstrates that the proposed MDE feature representation generalizes well across different turbine units and wind farms.

From an engineering perspective, these results indicate that the proposed method is not only effective under controlled experimental conditions but also exhibits strong robustness and transferability in practical deployment scenarios.



\section{Interpretability Analysis}

\subsection{Feature Space Visualization and Cross-Dataset Consistency}

\begin{figure}[htbp]
    \centering
    \includegraphics[width=1.0\linewidth]{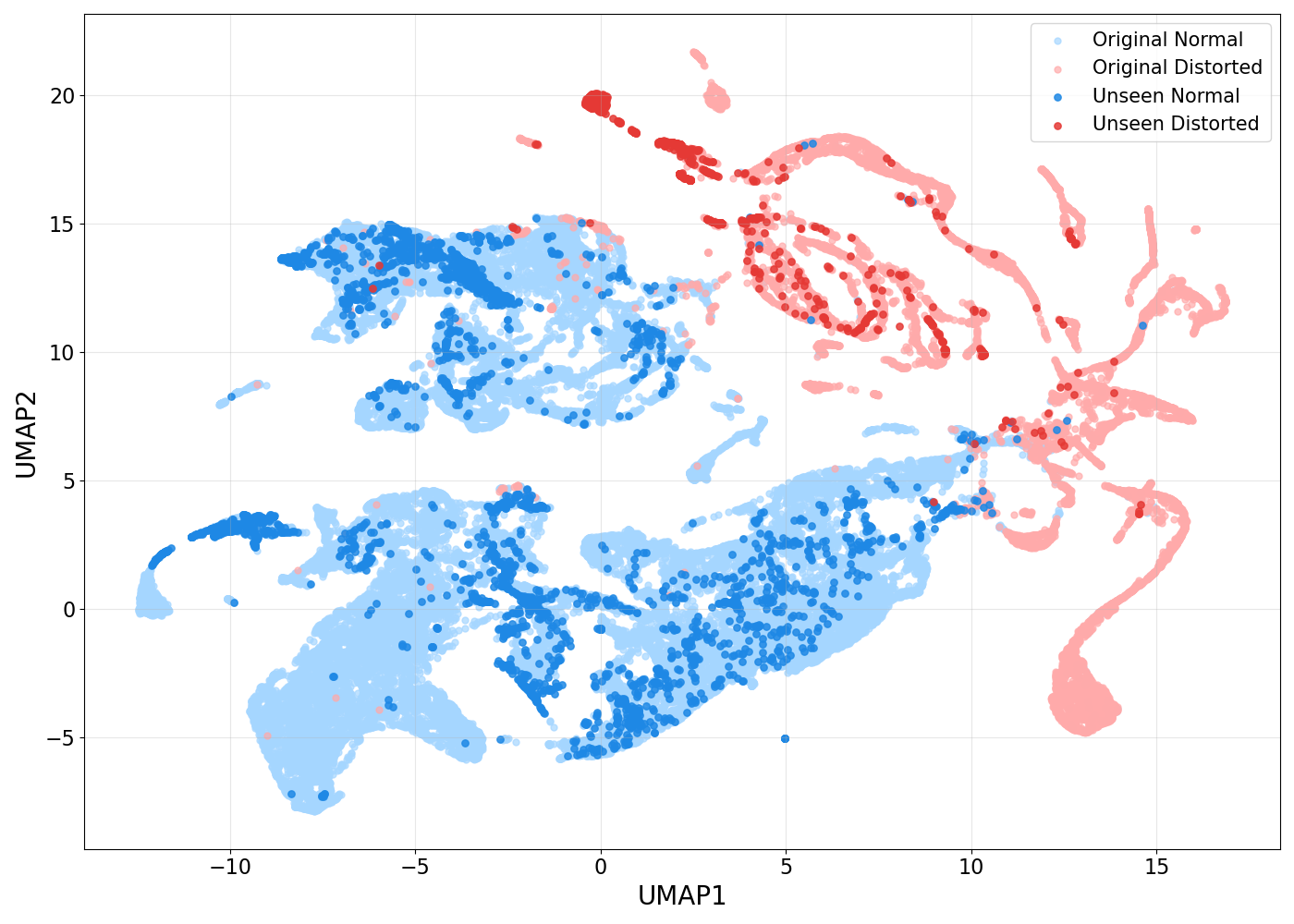}
    \caption{UMAP visualization of the seven MDE features. Normal samples are shown in blue tones (light: original dataset, dark: unseen dataset), and erroneous samples are shown in red tones (light: original dataset, dark: unseen dataset). The consistent clustering pattern across both datasets provides intuitive evidence for the discriminative structure and cross-dataset stability of the proposed MDE features.}
    \label{fig:umap_mde}
\end{figure}

Building on the within-dataset visualization reported in the initial study~\cite{mde1}, the seven-dimensional MDE feature space is jointly projected for the original and unseen datasets to examine cross-dataset consistency. Uniform Manifold Approximation and Projection (UMAP) is used for the two-dimensional embedding~\cite{u1}, as shown in Fig.~\ref{fig:umap_mde}.

Normal samples are shown in blue tones, with light blue representing the original dataset and dark blue representing the unseen dataset. Erroneous samples are shown in red tones, with light red representing the original dataset and dark red representing the unseen dataset. In the original dataset, normal and erroneous samples form largely separable clusters with only limited overlap. More importantly, the unseen dataset exhibits a similar separation pattern, despite potential distribution shifts caused by different turbine units, operating conditions, and acquisition settings.

This cross-dataset consistency suggests that the MDE features capture intrinsic characteristics of erroneous vibration signals rather than merely fitting dataset-specific patterns. The visualization therefore provides intuitive support for the quantitative results in Section~\ref{performance_across_model} and Section~\ref{sec:unseen_data_test}, where MDE shows stable model-agnostic performance and strong cross-turbine generalization.

\subsection{Statistical Distribution of Representative MDE Features}
To further examine the interpretability of the proposed MDE representation, the statistical distributions of representative features are analyzed using violin plots with embedded box plots~\cite{u2,u3}. To avoid overloading the visualization, one representative feature is selected from each MDE dimension. Specifically, RMS, TAE, SpAE$_1$, and FBEE$_2$ are selected to represent the energy, time-domain entropy, spectral entropy, and frequency-band energy entropy dimensions, respectively. This visualization is intended to provide distribution-level interpretation, while the complete feature-level and dimension-level quantitative comparisons have been reported in Section~\ref{robustness_analysis}.

\begin{figure}[htbp]
    \centering
    \includegraphics[width=1.0\linewidth]{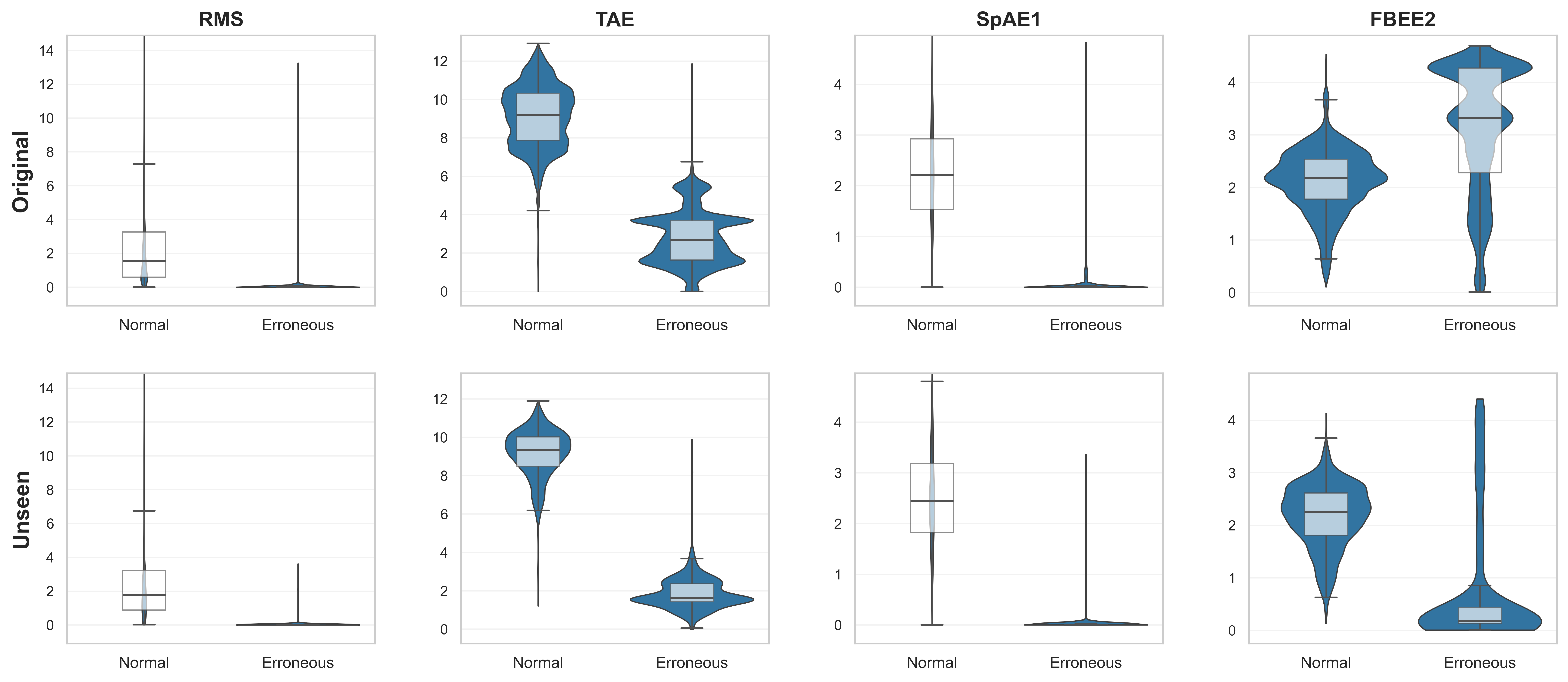}
    \caption{Statistical distributions of representative MDE features in the original and unseen datasets. 
    Each subplot shows the distribution difference between normal and erroneous samples using violin plots with embedded box plots. The consistent distribution patterns across the original and unseen datasets indicate that the statistical behavior of MDE features remains stable under cross-turbine data variations.}
    \label{fig:violin_and_box_mde}
\end{figure}

As shown in Fig.~\ref{fig:violin_and_box_mde}, normal and erroneous samples exhibit clear distribution differences in most representative features, reflected by separated medians, different dispersion ranges, and distinct distribution shapes. Similar distribution patterns are observed in both the original and unseen datasets, indicating that the statistical behavior of the MDE features remains stable across different turbines and acquisition conditions.

These observations support the cross-dataset consistency shown in the UMAP visualization and provide individual-feature-level evidence for the interpretability of MDE.

\subsection{Physical Interpretation of Entropy Variations}

The interpretability of MDE stems from its multi-perspective description of signal quality. Instead of relying on an abstract black-box representation, MDE characterizes erroneous vibration signals from four physically meaningful aspects: amplitude level, time-domain amplitude distribution, spectral organization, and frequency-band energy allocation.

Specifically, RMS represents the energy dimension and reflects the global amplitude level of the vibration signal. Abnormal energy variations may be induced by sensor malfunction, signal attenuation, shutdown transients, or abnormal acquisition conditions. TAE and ATAE describe the time-domain entropy dimension by quantifying the distribution characteristics of waveform amplitudes. SpAE1 and SpAE2 characterize the spectral entropy dimension by measuring how spectral amplitudes are distributed across frequency components. Erroneous signals may exhibit abnormal spectral flattening, concentration, or redistribution compared with structured vibration spectra under normal operating conditions. FBEE1 and FBEE2 further describe the frequency-band energy entropy dimension, providing complementary information on localized energy redistribution across predefined bands.

Therefore, the entropy variations observed in MDE correspond to distinct manifestations of signal degradation in amplitude level, time-domain waveform structure, spectral distribution, and band-wise energy allocation. This physically meaningful multi-dimensional design explains why MDE achieves both discriminative performance and robustness under heterogeneous industrial monitoring conditions.


\section{Discussion}

\subsection{Deployment Trade-off Across Layers}
\label{sec:deployment_tradeoff}

The cross-platform experiments confirm that the proposed MDE-based quality control pipeline can be deployed at different layers of a wind farm monitoring architecture. A centralized workstation is suitable for batch processing and integration with existing data-center infrastructure, whereas the N100 industrial PC and Jetson Orin Nano provide two feasible edge-side options.

The N100 IPC achieves the lowest per-signal latency (25.68--29.43 ms), benefiting from its x86 architecture and lightweight Linux environment. The Jetson Orin Nano requires a longer processing time (54.39--77.28 ms), but offers advantages in power efficiency and compact form factor, making it suitable for space- or power-constrained retrofit scenarios. In practical wind farm monitoring, vibration records are typically acquired intermittently at intervals of several hours per measurement point. Therefore, the observed processing latency is negligible relative to the acquisition interval and does not introduce a delay risk into the monitoring workflow. The deployment choice is thus mainly determined by infrastructure constraints, power budget, and integration cost, rather than by functional feasibility.

\subsection{Industrial Implications}
\label{sec:industrial_implications}

The proposed MDE framework addresses a practical but often overlooked problem in wind turbine condition monitoring: erroneous vibration signals may enter downstream diagnostic and prognostic models before any systematic quality control is performed. Such signals, caused by sensor malfunction, shutdown transients, or abnormal acquisition conditions, can lead to false alarms, missed detections, and unreliable maintenance decisions.

Embedding MDE as a front-end quality screening layer provides three practical benefits. First, sensor-related abnormal signals can be detected at an early stage, reducing the risk of using invalid measurements for machine health assessment. Second, automated screening reduces the need for manual signal inspection by domain experts, which is important for large-scale wind farm operation. Third, filtering erroneous signals before downstream analysis prevents error propagation into fault diagnosis and remaining useful life prediction, thereby improving the reliability of the overall monitoring pipeline.

Because MDE has low computational and memory requirements, it can be integrated into existing monitoring infrastructures as either a centralized preprocessing module or an edge-side screening layer, without requiring substantial additional computing resources.

\subsection{Limitations and Future Work}
\label{sec:limitations}
Several limitations should be acknowledged. First, the present evaluation is based on archived industrial data rather than live online deployment. Although the three tested hardware platforms represent realistic deployment targets, actual field performance may be affected by concurrent task scheduling, network latency, database access, and system integration overhead.

Second, the edge validation is limited to single-device deployment. Multi-node distributed deployment, where multiple edge devices coordinate with a central server, has not yet been evaluated.

Third, the dataset is restricted to semi-direct-drive turbine platforms in the 3 MW class from a single manufacturer. Although the dataset covers multiple wind farms, components, sensor brands, and sampling configurations, further validation is required for larger semi-direct-drive turbines, direct-drive turbines, DFIG-based turbines, and turbines from other manufacturers.

Future work will focus on live wind farm deployment, extension to other sensor modalities such as temperature, current, and acoustic emission signals, and integration with online learning mechanisms to adapt to long-term changes in industrial monitoring data.

\section{Conclusion}
\label{sec:conclusion}

This paper presents a systematic analysis of Multi-Dimensional Entropy (MDE) for vibration data quality control in wind turbines, with emphasis on computational efficiency, model-agnostic capability, physical interpretability, robustness, and deployment feasibility.

Experiments on more than 57,000 labeled industrial vibration samples from 12 wind farms demonstrate that MDE achieves stable performance across seven classifiers from different modeling paradigms, with ACC above 98.7\% and AUC above 0.99. Ablation studies show that the multi-dimensional design provides complementary and redundant signal representations, enabling robust performance even under reduced feature conditions. Cross-turbine validation further confirms that MDE maintains strong discriminative capability on unseen turbine units without retraining or fine-tuning.

Cross-platform deployment experiments show that the complete pipeline requires only 25--77 ms per vibration record, with peak memory increment below 5.5 MB. These results indicate that MDE can serve as a lightweight, interpretable, and deployment-ready feature layer for industrial vibration data quality control. By filtering erroneous signals before downstream fault diagnosis and remaining useful life prediction, MDE improves the reliability of wind turbine monitoring pipelines and supports more trustworthy maintenance decision-making.





\end{document}